\documentclass[runningheads]{llncs}

\usepackage{graphicx}
\usepackage{booktabs}
\usepackage{amsmath}
\usepackage{breakcites}
\usepackage{subfig,float,placeins}
\usepackage{hyperref}
\hypersetup{
  colorlinks=true,
  allcolors=blue
}

\begin{document}

\title{Texture Feature Analysis for Classification of Early-Stage Prostate Cancer in mpMRI
\thanks{Supported by Azzaytuna University and the Ministry of Higher Education and Scientific Research, Libya and Supercomputing Wales.}}
\titlerunning{Texture Feature Analysis for Classification of Early-Stage Prostate Cancer}

\author{Asmail Muftah\inst{1,2}\orcidID{0000-0003-4174-8595}
\and
SM Shermer\inst{3}\orcidID{0000-0002-5530-7750}
\and
Frank C Langbein\inst{1}\orcidID{0000-0002-3379-0323} 
}
\authorrunning{A.\ Muftah et al.}

\institute{School of Computer Science and Informatics, Cardiff University, Cardiff, UK\\
\email{MuftahA@cardiff.ac.uk}, \email{LangbeinFC@cardiff.ac.uk}
\and
School of Science, Azzaytuna University, Tarhounah, Lybia\\
\email{asmail.muftah@azu.edu.ly}
\and
Faculty of Science \& Engineering (Physics), Swansea University, Singleton Park Campus, Swansea, UK\\
\email{s.m.shermer@gmail.com}
}

\maketitle

\begin{abstract}
Magnetic resonance imaging (MRI) has become a crucial tool in the diagnosis and staging of prostate cancer, owing to its superior tissue contrast. However, it also creates large volumes of data that must be assessed by trained experts, a time-consuming and laborious task. This has prompted the development of machine learning tools for the automation of Prostate cancer (PCa) risk classification based on multiple MRI modalities (T2W, ADC, and high-b-value DWI). Understanding and interpreting the predictions made by the models, however, remains a challenge. We analyze Random Forests (RF) and Support Vector Machines (SVM), for two complementary datasets, the public Prostate-X dataset, and an in-house, mostly early-stage PCa dataset to elucidate the contributions made by first-order statistical features, Haralick texture features, and local binary patterns to the classification. Using correlation analysis and Shapley impact scores, we find that many of the features typically used are strongly correlated, and that the majority of features have negligible impact on the classification. We identify a small set of features that determine the classification outcome, which may aid the development of explainable AI approaches.
\end{abstract}

\section{Introduction}

The early detection of prostate cancer (PCa) typically relies on blood tests such as the prostate-specific antigen (PSA) test and digital rectal examination (DRE), followed by transrectal ultrasound (TRUS) biopsy~\cite{smith2016cancer}. However, TRUS biopsy carries the risk of serious complications, including meningitis and sepsis~\cite{loeb2013systematic}. Furthermore, it can lead to the detection of clinically insignificant or indolent cancer, resulting in overdiagnosis and potentially unnecessary treatments~\cite{abraham2015patterns}. To improve diagnosis and minimize unnecessary biopsies, multiparametric MRI (mpMRI) has become the standard of care in the diagnosis and staging of PCa. The resulting high volume of imaging data generated has in turn stimulated significant research efforts to develop effective machine learning tools to assist radiologists with the segmentation and classification of lesions, and both traditional and deep-learning methods have been applied successfully to the problem of identifying clinically significant lesions (see, e.g.,~\cite{aldoj2020semi,chen2019transfer,deniffel2020using,fehr2015automatic,litjens2014computer,liu2009prostate,niaf2011computer,vos2008computerized,yuan2019prostate,zhong2019deep}).

Despite the increasing popularity of deep learning approaches, recent work evaluating the performance of convolutional neural networks (CNNs) and transfer learning, as well as traditional machine learning classifiers based on handcrafted features such as first-order statistics, Haralick features~\cite{haralick1973textural} and local binary patterns (LBP)~\cite{1017623}, suggests that traditional machine learning methods such as Support Vector Machines (SVM) and especially Random Forest (RF) classifiers can perform at least as well as deep learning classifiers or better~\cite{cameron2015maps,chen2019prostate,fehr2015automatic,Muftah2023,liu2019prediction,orczyk2019prostate,peng2013quantitative,wu2019transition}. Machine learning tools based on handcrafted features are also well suited to feature analysis with the aim of explainability. Explainability remains a challenge for machine learning, but recent work, for example, has explored explanable AI to predict cancer based on gene expression~\cite{RAMIREZMENA2023107719}, and survival rates predicted by synoptic reporting of pathology~\cite{janssen2022using}.

In this work, we study explanability in the context of PCa classification based on mpMRI data by exploring the features used by the best-performing traditional machine learning classifiers to understand which are most relevant, and their respective impact on the classification results. The best-performing classifiers trained for classifying rectangular prostate patches into suspicious (positive) and normal (negative) based on first-order statistical features, Haralick texture features and LBP using sequential backward floating feature selection (SBSF) are identified by clustering their performance according to multiple performance metrics (AUC, Accuracy, F1-score, sensitivity and specificity). All classifiers are trained on two complementary datasets, the public ProstateX database, and an in-house dataset of patients with suspected early-stage PCa, as well as a dataset combing these two. Feature correlation and utilization are studied using correlation analysis and Shapley impact scores to reveal a small set of features that consistently explain most of the classification results for both datasets.

\section{Datasets}

To assess the performance of various machine learning algorithms, two datasets are utilized: the publicly available ProstateX dataset~\cite{litjens2017prostatex} and an in-house collection of anonymized mpMRI data, primarily representing early-stage PCa. The ProstateX dataset comprises $194$ negative and $71$ positive samples. The in-house dataset comprises $44$ negative and $46$ positive samples, selected from a cohort of patients who had undergone mpMRI scans at a local clinical imaging unit. Leveraging both datasets enhances the comprehensiveness of our evaluation, provides valuable insights into the applicability of algorithms across diverse patient cohorts, and facilitates a deeper understanding of their robustness and generalizability in a broader clinical setting.

\begin{figure*}[t]\centering
\includegraphics[width=0.9\textwidth]{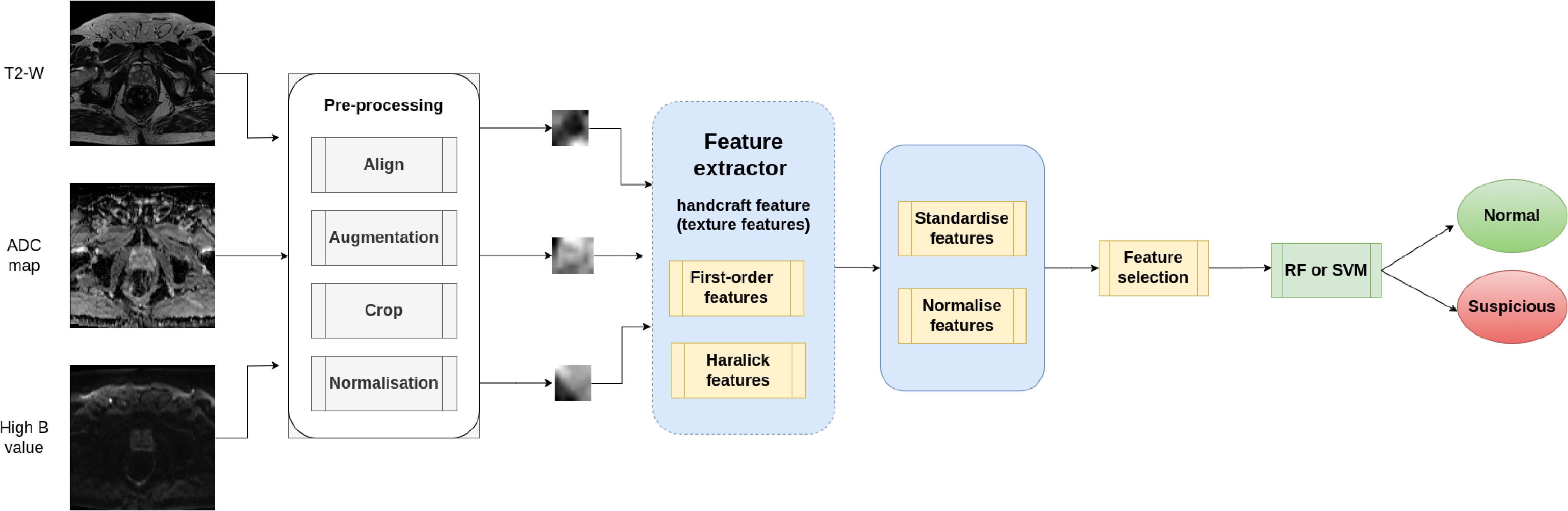}
\caption{Traditional machine learning classification pipeline.}\label{fig:ML-overview}
\end{figure*}

For both datasets, we incorporate (axial) T2-weighted (T2W) images, apparent diffusion coefficient (ADC) maps, and high-b-value diffusion-weighted images (DWI) generated by the imaging system, as illustrated in Fig.~\ref{fig:ML-overview}. All modalities are registered using the patient coordinate system from the DICOM files, as manual verification suggests that further automated registration is prone to introducing larger errors. The 12-bit intensity values in each modality are rescaled to the range $[0, 1]$. Given the limited size of the datasets, data augmentation techniques are employed to generate additional samples for each patient in the respective datasets, adding $39$ samples per patient. Augmentation methods include rotation, flipping, scaling, elastic deformation, shearing, Gaussian noise, blur, and adjustments to contrast and brightness. Subsequently, PCa lesions are extracted as 2D patches and resized to either $16\times 16$ or $32\times 32$ based on segmentation masks indicating negative and positive regions for PCa. For the in-house early-stage PCa dataset, classification relies on a set of normal and suspicious regions identified by the reporting radiologist. For the ProstateX dataset, lesion classification is based on the methodology outlined in~\cite{cuocolo2021quality}.

\section{Methods}

In recent work, many configurations for different classifier types were systematically investigated across various parameters for both datasets, including traditional machine learning with handcrafted features and deep learning, pre-trained or trained from scratch. Various configurations for each classifier type were investigated across a range of parameters, and each classifier configuration evaluated according to several standard metrics, including the area under the curve (AUC) of the receiver operating characteristic (ROC), accuracy, F1 score, sensitivity and specificity. In this study we only consider traditional machine learning results as they use explicit features suitable for explanability analysis. The code for the classifiers and the complete training and analysis results are available at~\cite{results-anon,code-ref-anon}.

First-order statistical features, Haralick texture features, and LBP were calculated for all three MRI modalities used (T2W, ADC, high-b-value DWI), and fed into sequential backward floating feature selection (SBFS) to identify well-performing features and eliminate redundant features, focusing on two of the most frequently used machine learning classifiers, SVM and RF, as illustrated in Fig.~\ref{fig:ML-overview}. This analysis explored different kernel functions and regularization options for SVM, and various hyperparameters for RF, including the number of individual decision trees composing the forest ($50$, $100$, or $150$), the maximum depth of each tree ($0$, $20$), minimum samples per leaf ($2$ or $4$) and split ($1$, $20$, or $40$) defining the prerequisites for further bifurcations of decision nodes in the tree, as well as preprocessing steps, including intensity standardization, normalization, and combinations of both. Five-fold cross-validation is used to evaluate the effectiveness of the machine learning models and training dataset dependency, resulting in five evaluation scores, reported by their mean and std.\ deviation across the folds. Based on the results the best traditional machine learning models for each dataset are selected for our feature analysis.

In this work clustering, taking into account all performance metrics, is used to identify the best-performing classifier configurations for further analysis, to understand which features are selected and their relative importance in the classification process. Linear correlation coefficients between feature vectors are calculated to understand the degree of independence of different features. This is especially important due to the large number of features involved and expected redundancy of certain features due strong statistical correlation between certain first-order statistical and texture features, for example.

To elucidate the impact of individual features on the classification results, Shapley values~\cite{10.5555/3295222.3295230}, quantifying the average marginal contribution of each feature value to the overall score across all possible combinations, are calculated for all features. This enables us to identify the set of features that contribute the most to the classification, as well as those whose contribution is negligible. In addition to facilitating understanding of how the classifiers make decisions, identification of subsets of relevant features that are consistently used for different datasets by the best-performing classifiers, could be leveraged to reduce the number of features that need to be calculated and develop more efficient algorithms. It may also help to drive approaches towards explainable AI to refer to specific textures and their relations between different modalities in regions of the prostate suspected of being cancerous or not, in traditional as well as deep machine learning.

\section{Results}

\subsection{Best-performing machine learning classifiers}

\begin{figure}[t]\centering
\subfloat[Prostate X dataset]{\includegraphics[width=0.47\textwidth]{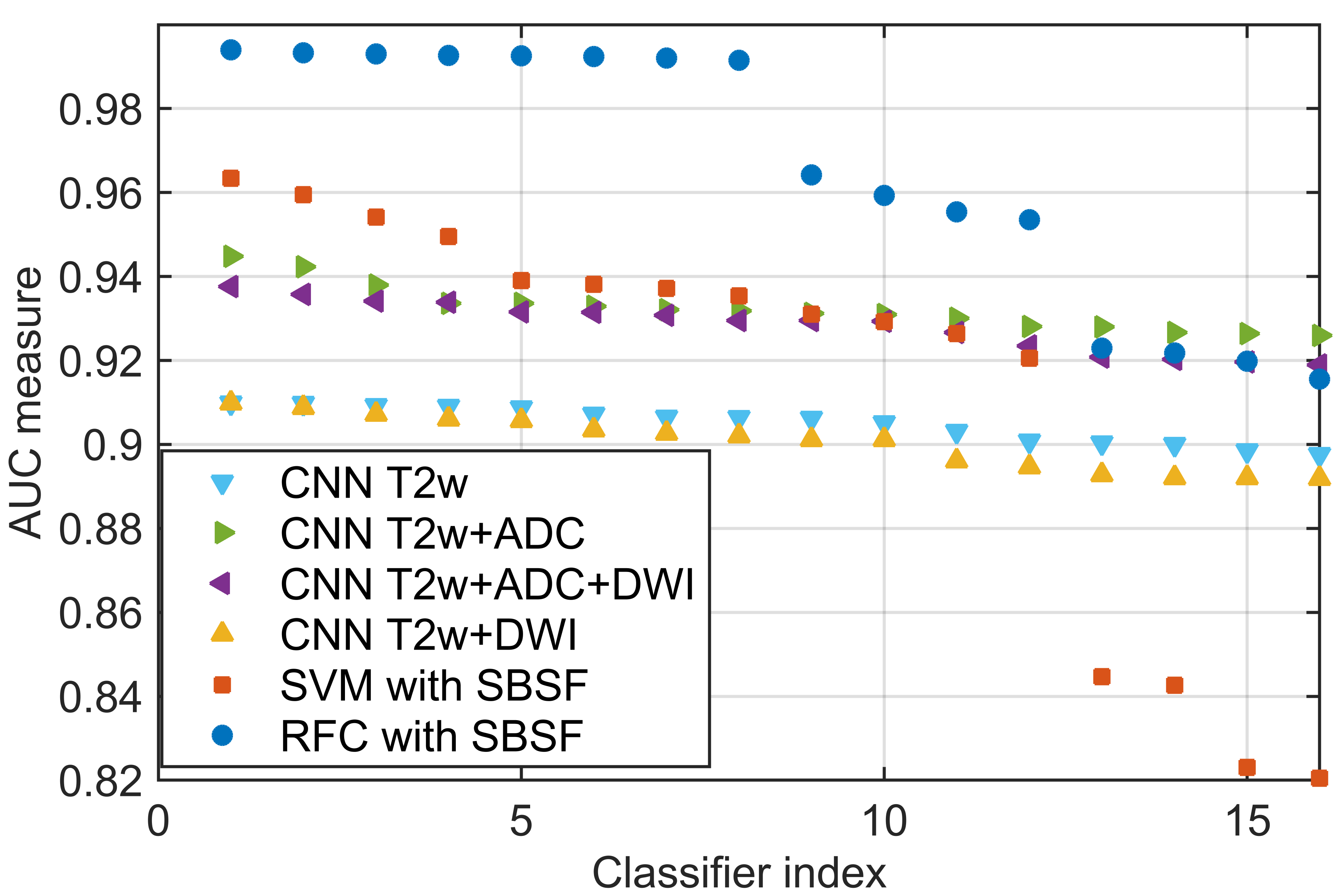}}
\hfill
\subfloat[In-house dataset]{\includegraphics[width=0.47\textwidth]{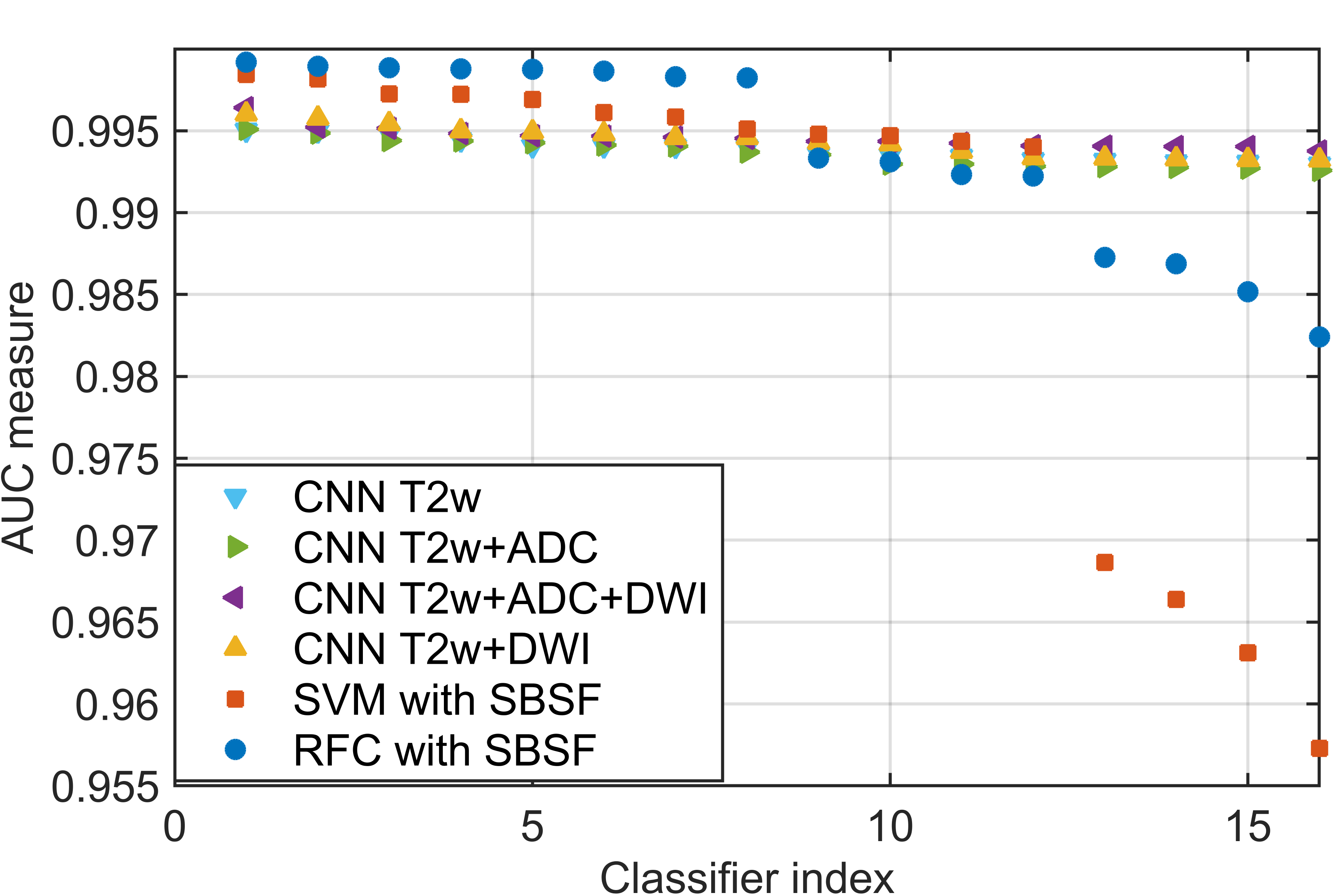}}
\caption{AUC metrics of the best classifiers for different types of traditional (RF and SVM with SBSF) and deep learning classifiers (CNNs trained on different combinations of inputs: T2W, T2W+ADC, T2W+ADC+DWI, T2W+DWI) for Prostate-X and in-house datasets show that traditional machine learning classifiers of RF type consistently perform best.}\label{fig:AUC-all}
\subfloat[Prostate X]{\includegraphics[width=0.47\textwidth]{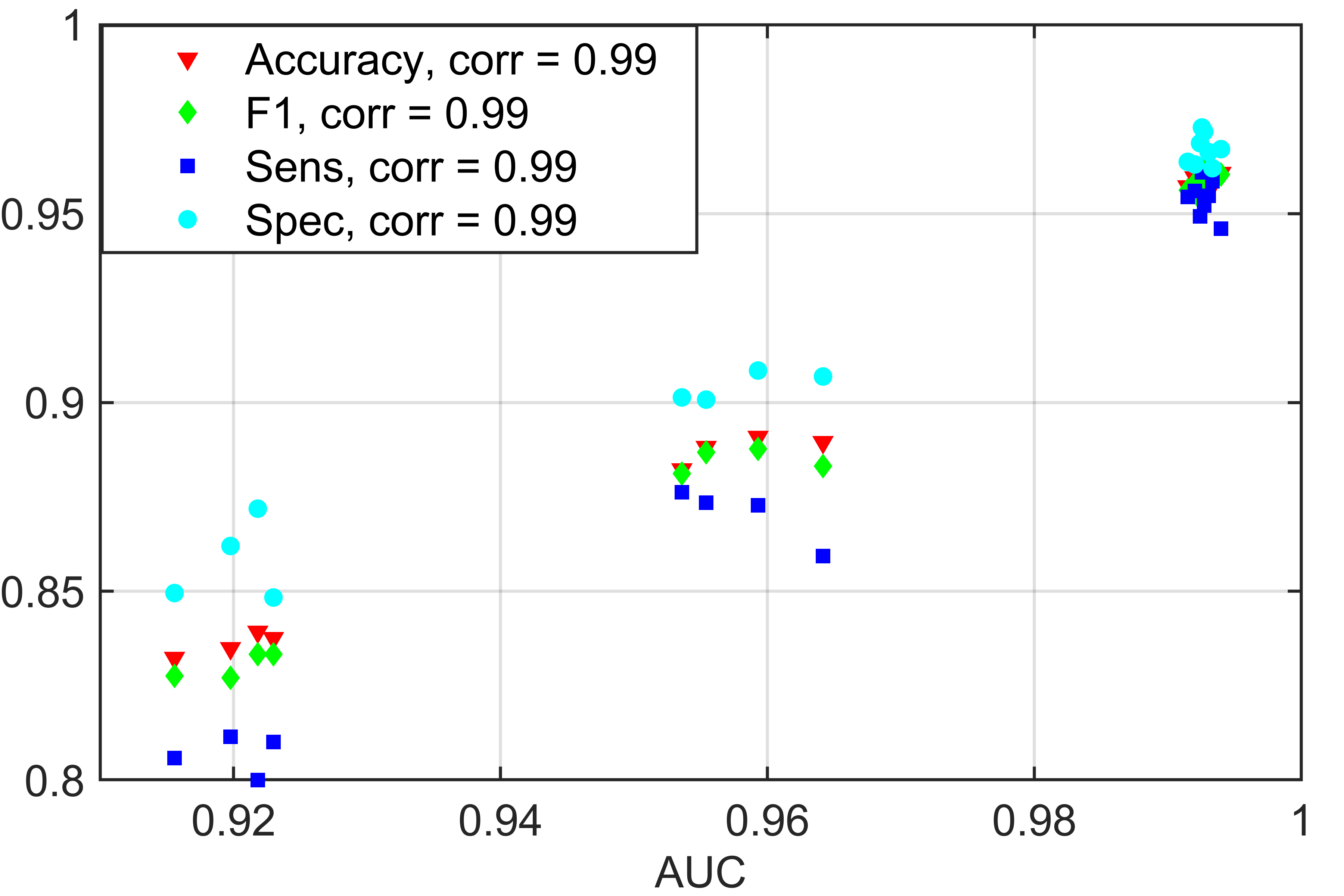}} \hfill
\subfloat[In-house data]{\includegraphics[width=0.47\textwidth]{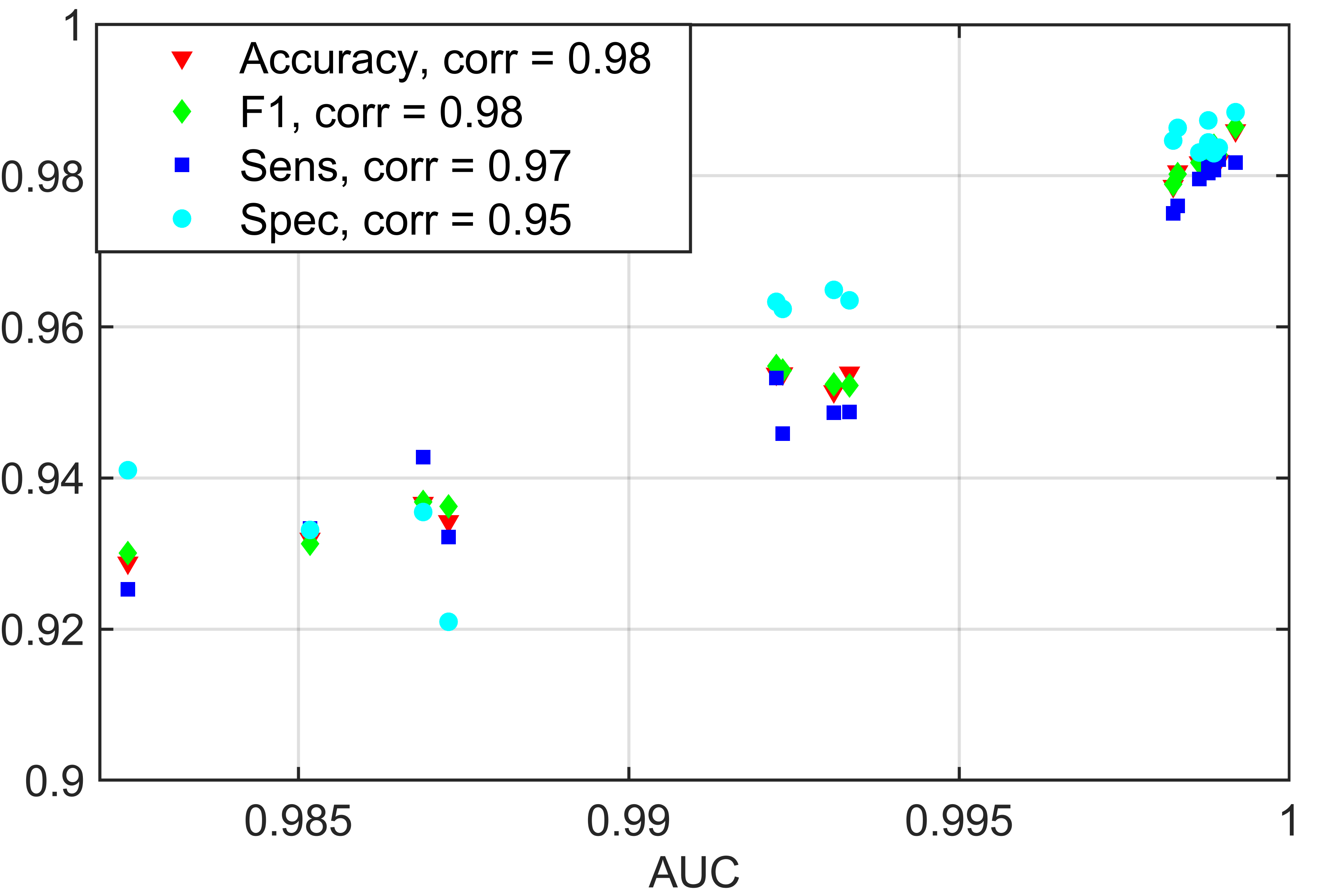}}
\caption{Accuracy, F1-score, sensitivity and specificity are strongly correlated with AUC, supporting use of AUC as the primary metric for ranking classifiers.}\label{fig:Perf-metrics}
\subfloat[Prostate X]{\includegraphics[width=0.49\textwidth]{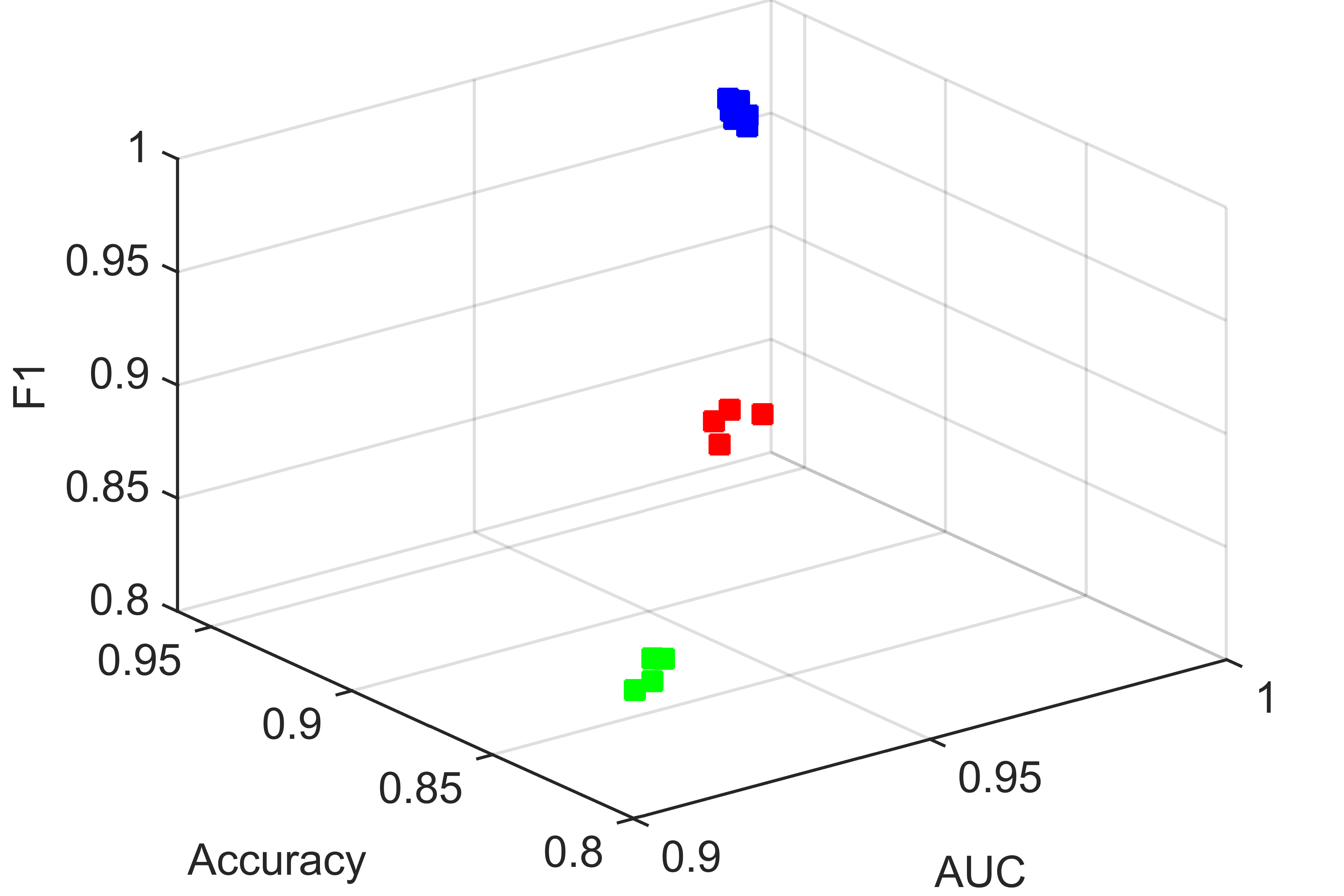}} \hfill
\subfloat[In-house data]{\includegraphics[width=0.49\textwidth]{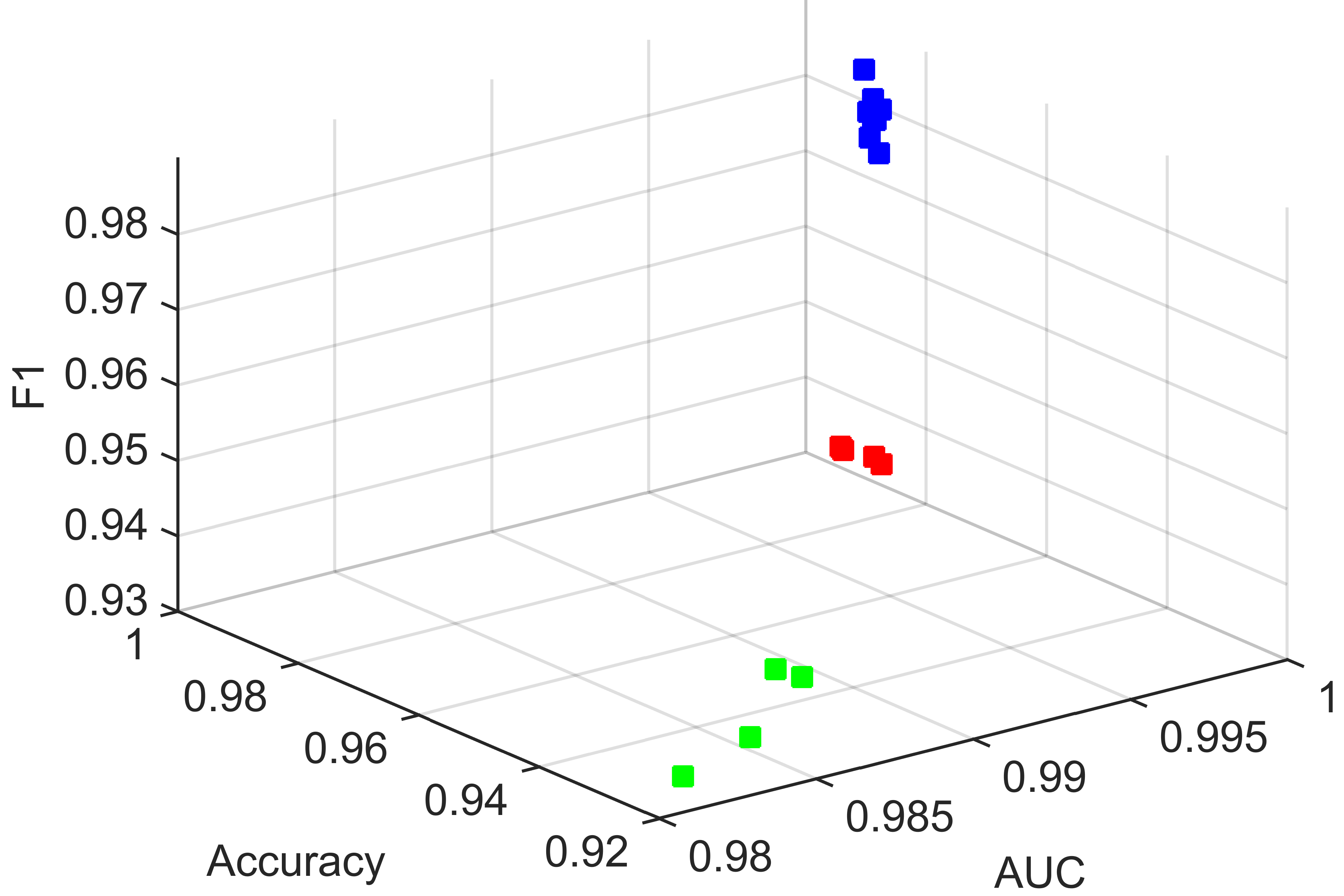}}
\caption{RF classifiers fall into three distinct performance clusters when clustered according to AUC values, accuracy, F1-score, sensitivity and specificity for both datasets (sensitivity and specificity dimensions not shown). The elements in the cluster with the highest performance scores (indicated by blue squares) correspond to identical configurations for both datasets.} \label{fig:Perf-clusters} 
\end{figure}

Fig.~\ref{fig:AUC-all} shows that the best-performing classifiers for both datasets are traditional machine learning classifiers of RF-type when ranked according to AUC. The same also holds for the combined dataset (not shown).

\FloatBarrier

Fig.~\ref{fig:Perf-metrics} further shows that the performance metrics for the best-performing classifiers for different datasets are strongly correlated with the AUC metric, justifying use of the former as a primary performance indicator for ranking different classifiers. The figures include results from CNN deep learning classifiers for reference and comparison, which are not further studied here as we wish to identify relevant explicit texture features.

Clustering of the best-performing classifiers by mean AUC, accuracy, F1-score, sensitivity, and specificity shows that, for both the Prostate-X and the in-house dataset, three performance clusters are identified, as shown in Fig.~\ref{fig:Perf-clusters}. The top-performing cluster for both datasets correspond to identical configurations for the classifiers comprising RF classifiers with $100$ trees, no maximum depth, two minimum samples per leaf and one split as prerequisite for further bifurcation. Both patch sizes of $16$ and $32$ as well as combinations of normalization, standardization or both, are included. It should be noted that the classifiers, although having the same configurations, were trained independently per dataset.

\begin{figure}[t]
\centering
\includegraphics[width=1\textwidth]{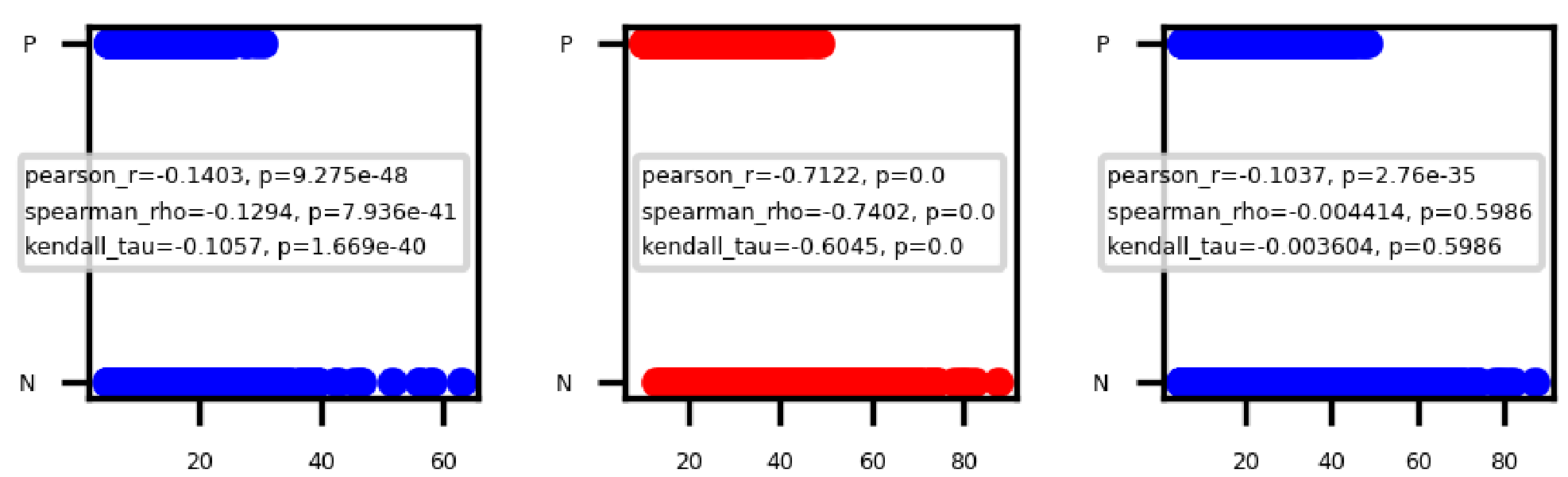}
\caption{Feature value ranges for positive and negative patches differ but generally overlap for all datasets, as illustrated here for the mean values of the T2W patches. Left: Prostate-X, middle: in-house dataset, right: combined dataset. Red color indicates strong correlation in terms of Pearson $r$.}\label{fig:feature-ranges}
\end{figure}

\begin{figure}[!t]
\includegraphics[width=1\textwidth]{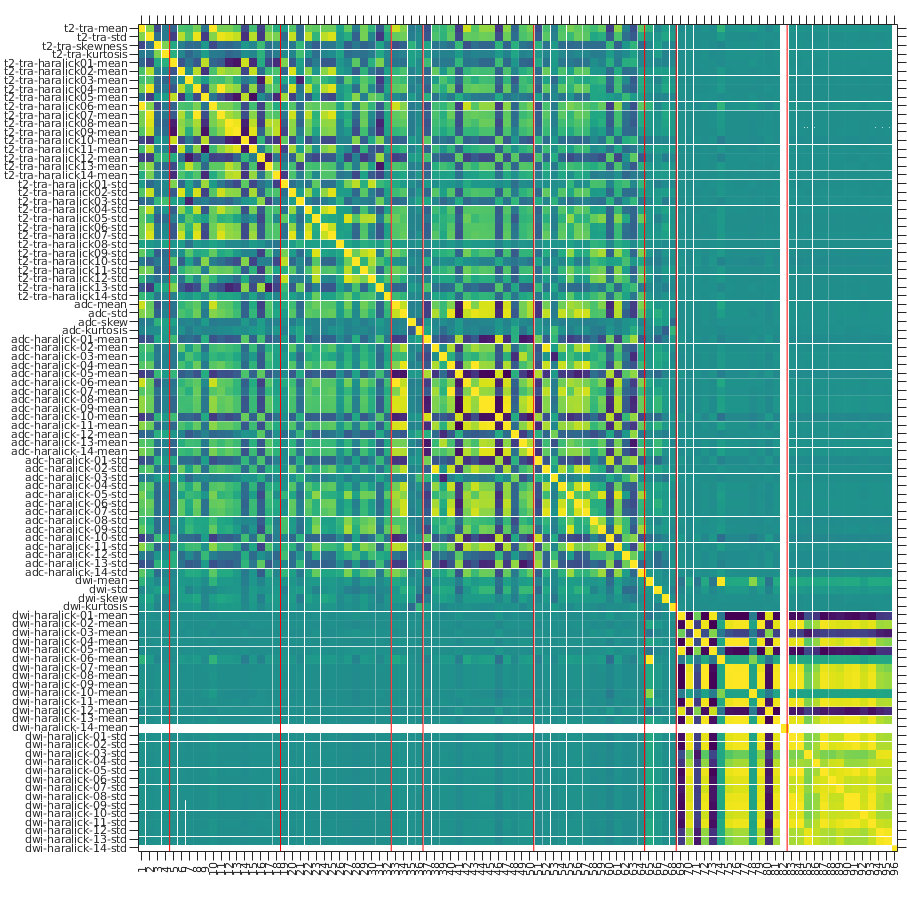}
\caption{Feature value correlation matrix for the Prostate-X dataset. The strength of correlation is indicated by the color, with yellow indicating high correlation and blue effectively none. The plot shows that many features are correlated but there are groups of uncorrelated features. The white line corresponds to a feature (for high-b-value DWI) that could not be computed.}\label{fig:features-correlations}
\end{figure}

\subsection{Feature value ranges and correlations}

We study the distributions of feature values for the positive and negative classes. Features are labelled by modality (t2-tra, adc, dwi\_c-1400) followed by feature type (mean, std, skewness, kurtosis, Haralick01 to Haralick14, lbp-01 to lbp-35) thoughout.  Fig.~\ref{fig:feature-ranges} shows the distribution of values for an illustrative example. The feature value ranges for both classes overlap for all datasets, but the ranges are narrower for the positive class, and for this feature, the positive class values are consistently on the lower side. A narrower range of values for the positive class, yet considerable overlap with the negative class, is typical for most features, showing that no single feature is sufficient for classification, as expected. For the other features generally similar behaviour is observed.

Given the large number of features across three different modalities, we expect many features to be correlated. To understand the degree of correlation between features, Fig.~\ref{fig:features-correlations} shows the feature correlation matrix for the Prostate-X dataset. It can be observed that the first-order statistical means are almost perfectly correlated with the Haralick06 feature values, and there is significant correlation between mean and standard deviation among the first-order order statistical features for the T2W images, while skewness and kurtosis are generally uncorrelated with the other first-order statistical features and most Haralick features. Therefore, it is expected that not all features contribute equally to the classification, and some may be entirely superfluous. The aim of feature selection, as applied via SBFS in training the classifiers, is to eliminate these correlations by removing features that do not significantly change the performance.

\subsection{Feature impact on model output: Shapley values}

\begin{figure}[!t]
\centering
\subfloat[Prostate-X]{\includegraphics[height=3in]{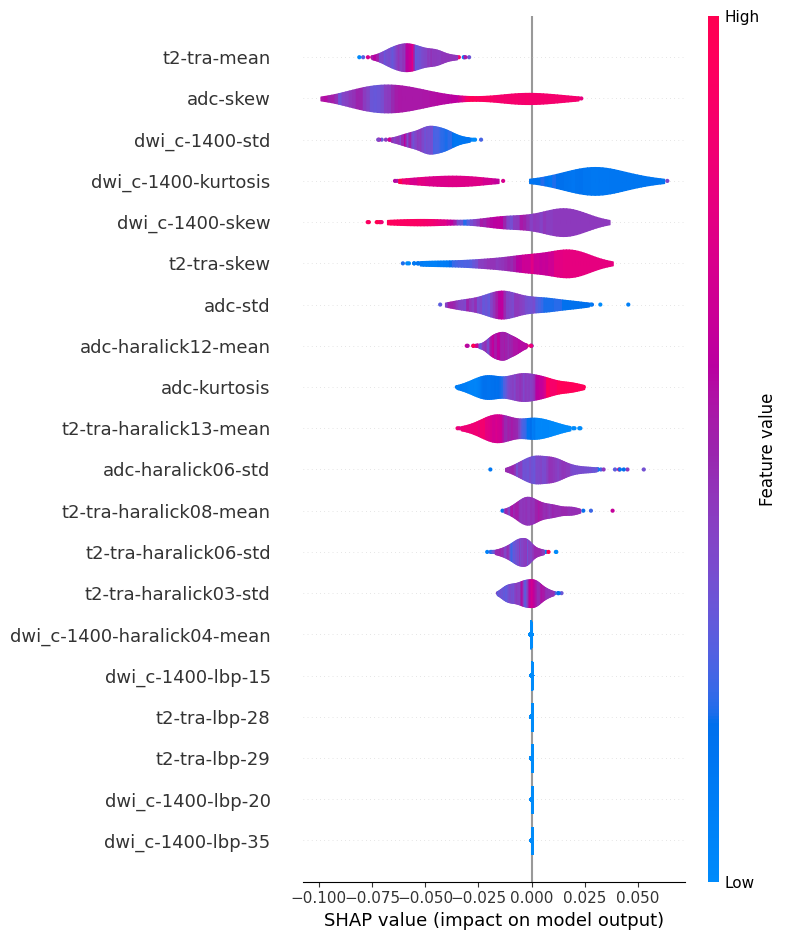}} \qquad
\subfloat[In-house dataset]{\includegraphics[height=3in]{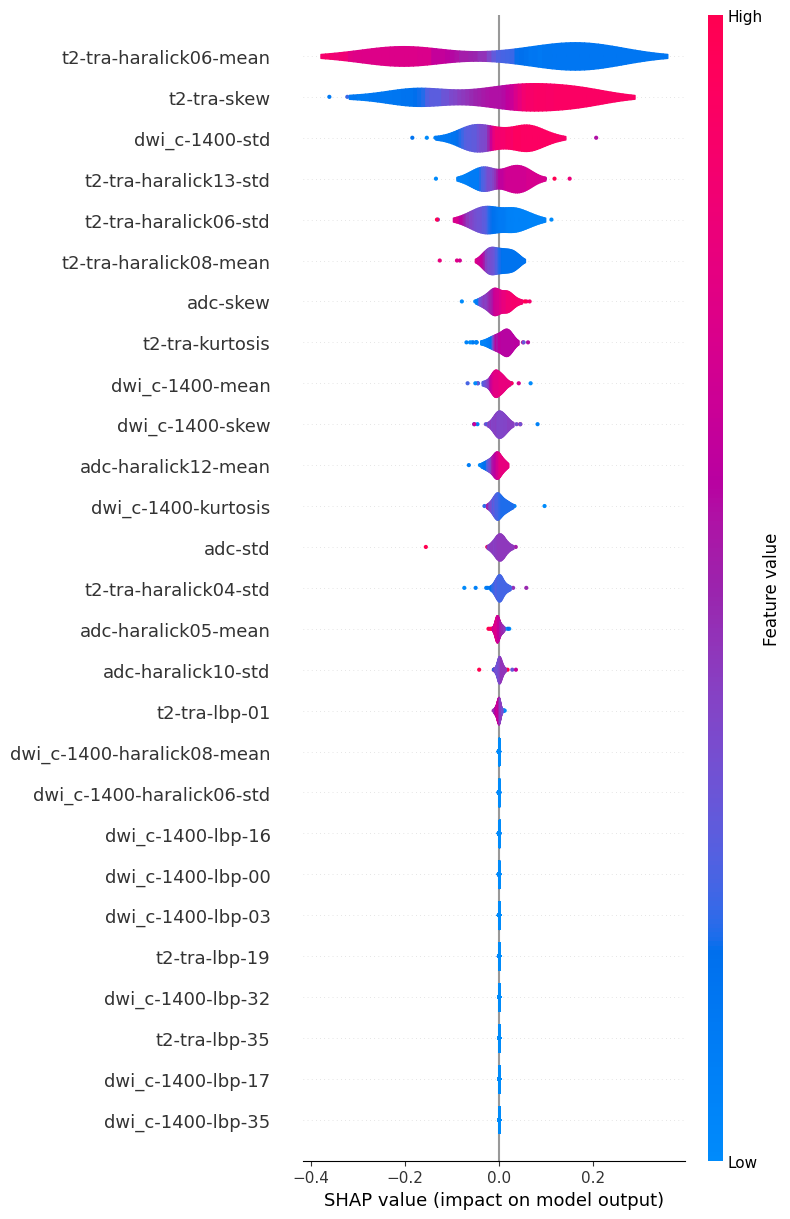}}\\
\subfloat[Prostate-X]{\includegraphics[width=0.49\textwidth]{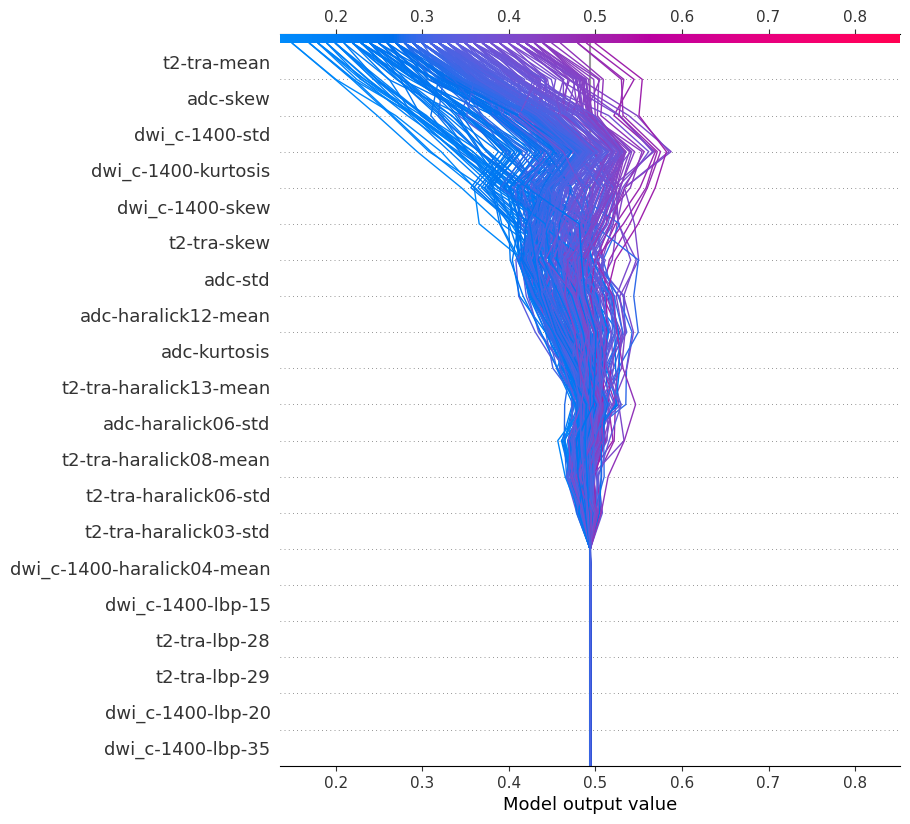}}
\subfloat[In-house dataset]{\includegraphics[width=0.49\textwidth]{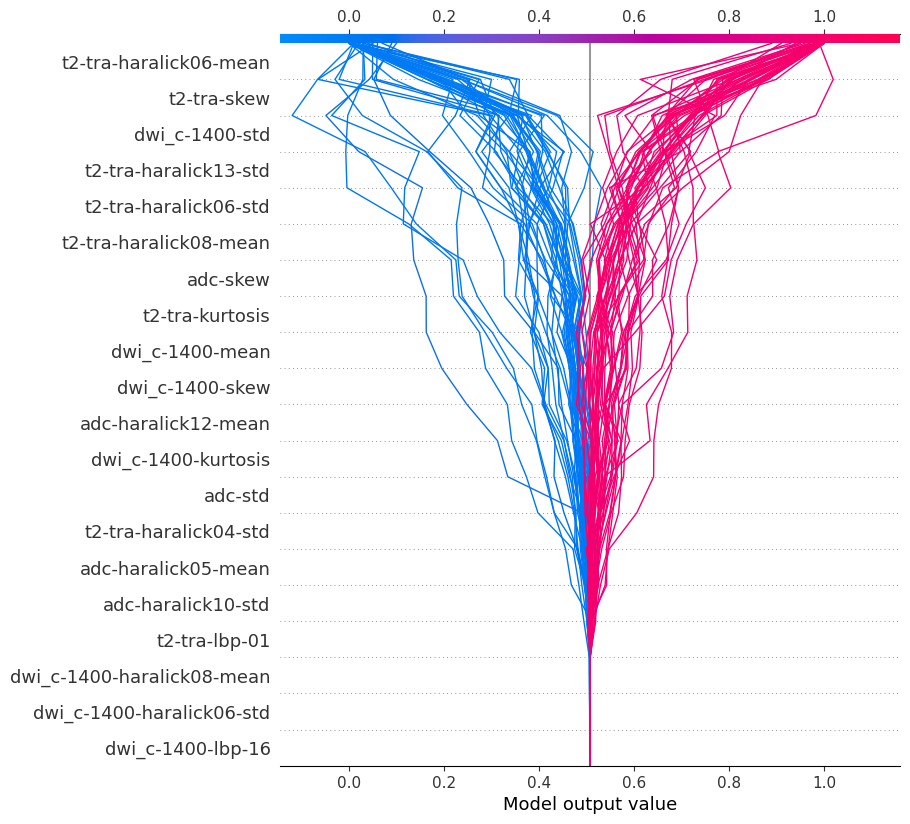}}
\caption{Shapley value distributions and decision trees for the best RF classifier with SBSF feature selection for the respective datasets.}
\label{fig:Shapley1}
\end{figure}

To gain a better understanding of the impact of individual features on the model output, we calculate the distribution of Shapley values for the features selected by SBFS for the best performing RF model for both datasets.  SHAP (SHapley Addictive exPlanations) was chosen over LIME as it is generally more suitable for complex machine learning models. A detailed description and discussion of the methods can be found, e.g., in~\cite{Sathyan2022}.  The violin and decision tree plots (Fig.~\ref{fig:Shapley1}) show the distribution of the Shapley values for each feature, with color coding indicating the feature values. The plots show that in both cases the model output is almost entirely determined by a small number of around $14$ features. For the ProstateX data, $8$ of the $14$ features that have a non-negligible impact on the model output are first-order statistical features; the remaining six are Haralick texture features ($12$, $13$, $06$, $08$, $06$, $03$); no LBP features were used. First-order statistical features clearly dominate, especially T2W mean, as well as skewness (used for all modalities), kurtosis (ADC, DWI) and std (ADC, DWI). It is also noteworthy that features from all three modalities are used ($6$ T2W, $5$ ADC, and $3$ DWI). Although fewer DWI features are used, their impact on the overall model output is significant.

For the in-house dataset comprised of mostly early-stage cancer patients, although the shape of the Shapley distributions and the impact of the individual selected features differ, the overall picture in terms of the relevant features is broadly similar, with perhaps a slightly higher contribution of Haralick texture features -- about half the features used are first-order statistical, and half are Haralick texture ($06$, $13$, $06$, $12$, $04$, $05$, $10$) features. Again features from all three modalities are used ($6$ T2W, $5$ ADC, $4$ DWI), with skewness and kurtosis playing significant roles in addition to mean and standard deviation (std). It could be argued that one of the LBP features contributes marginally but again, LBP feastures do not appear to play a significant role.

\begin{figure}[t]\centering
\subfloat[]{\includegraphics[width=0.49\textwidth,height=3in]{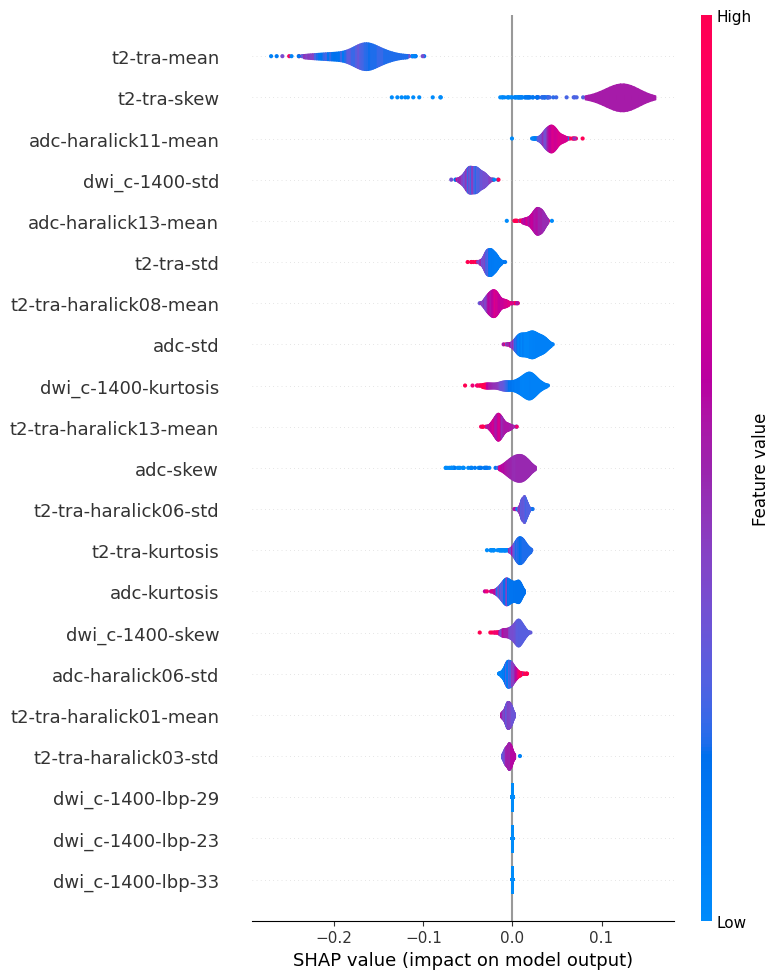}}\hfill
\subfloat[]{\includegraphics[width=0.49\textwidth,height=3in]{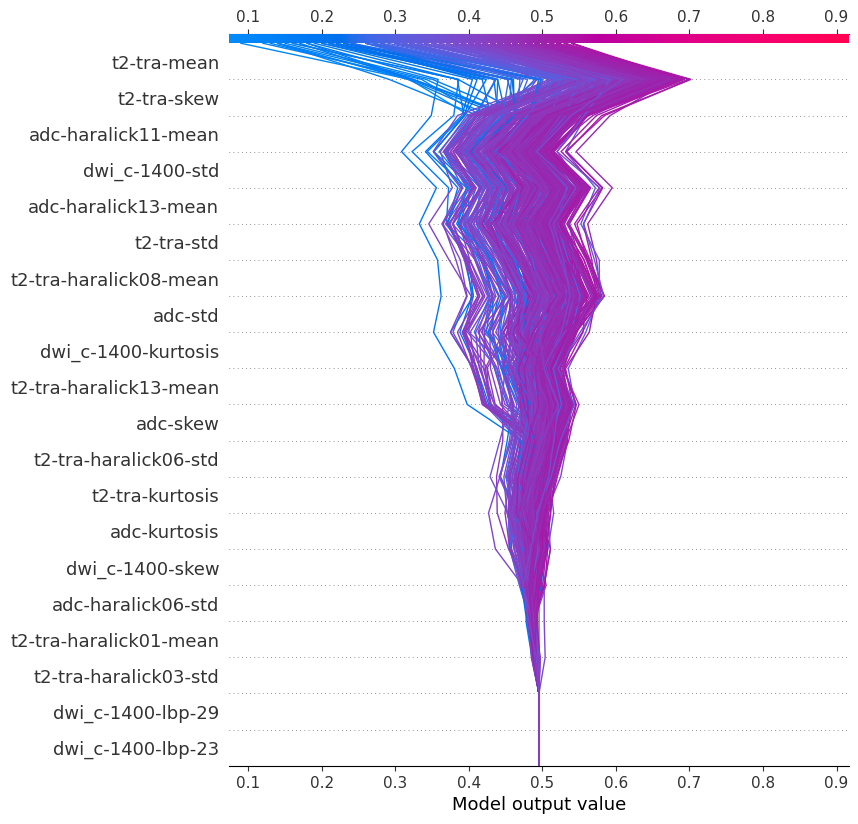}}
\caption{Violin plots of Shapley value distributions for the combined dataset.}\label{fig:Shapley-combi}
\end{figure}

We also considered the performance of classifiers trained on the combined (Prosate-X + in-house) dataset. The Shapley values distributions for the features, shown in Fig.~\ref{fig:Shapley-combi}, again show that only a few features impact the final result, again dominated by first-order statistical features ($10$ out of $18$) and a select number of Haralick texture features ($11$, $13$, $08$, $13$, $06$, $01$, $03$) with negligible contributions from LBPs. Features from all three modalities are used, with T2W mean as well as skewness and kurtosis featuring prominently, and std also being used.

One interesting difference between the Prostate-X (and combined) dataset vs the in-house dataset is that the decision paths are much clearer for the in-house dataset, although part of this could simply be that the dataset is smaller and thus lacks the full co-variance and range of feature values present.

\subsection{RF classifiers trained without feature selection}

\begin{figure}[!t]
\subfloat[ProstateX dataset]{\includegraphics[width=0.49\textwidth]{figures/output_6_3.png}} \hfill
\subfloat[In-house dataset]{\includegraphics[width=0.49\textwidth]{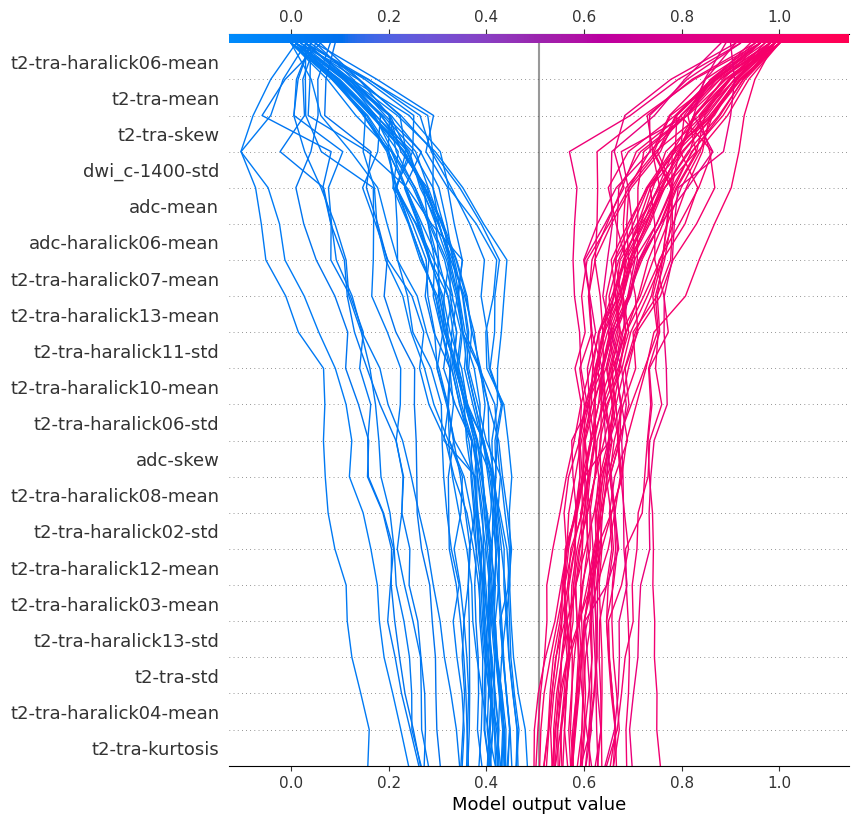}}
\caption{Shapley value analysis showing the most significant features used by best RF classifiers, trained using all features without feature selection. The best classifier for both datasets defaulted to an RF classifier with $100$ decision trees, no maximum depth, minimum of two samples per leaf.}\label{fig:rfc-shapley-no-feature-selection}
\end{figure}

\begin{figure}
\subfloat[Prostate-X\label{fig:rfc-shapley-no-feature-selection-px}]{\includegraphics[width=0.49\textwidth]{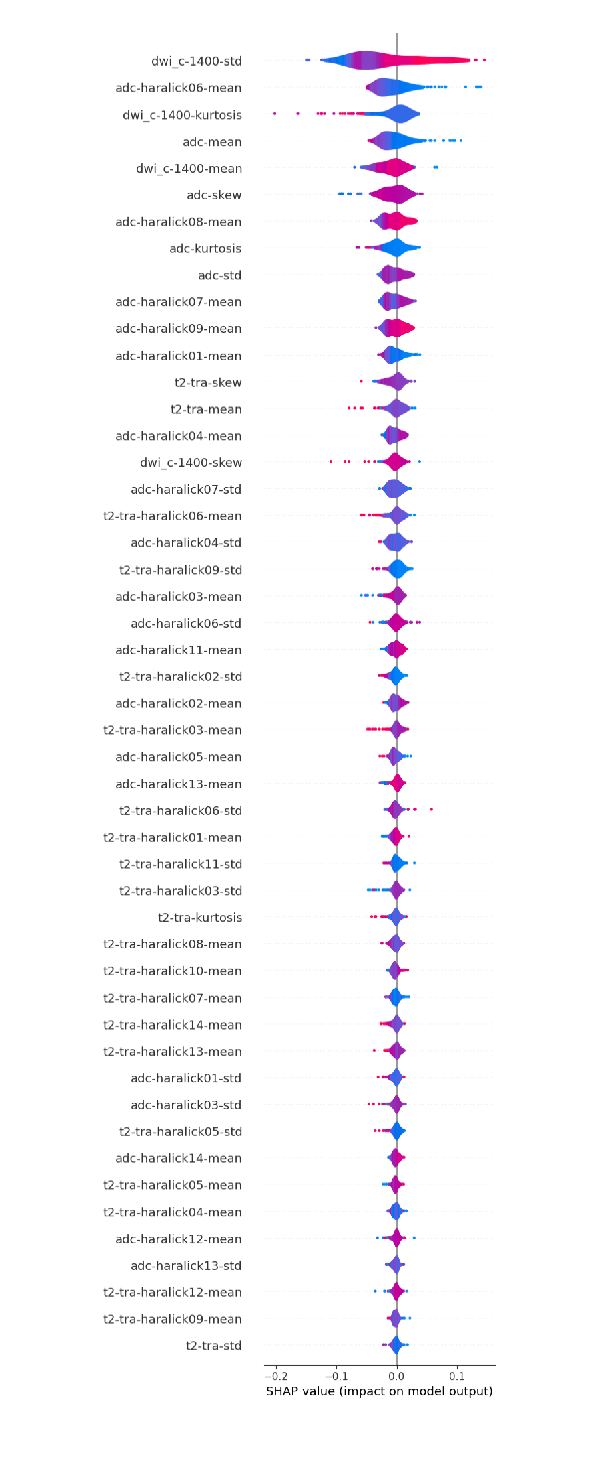}}
\subfloat[In-house dataset\label{fig:rfc-shapley-no-feature-selection-sw}]{\includegraphics[width=0.49\textwidth]{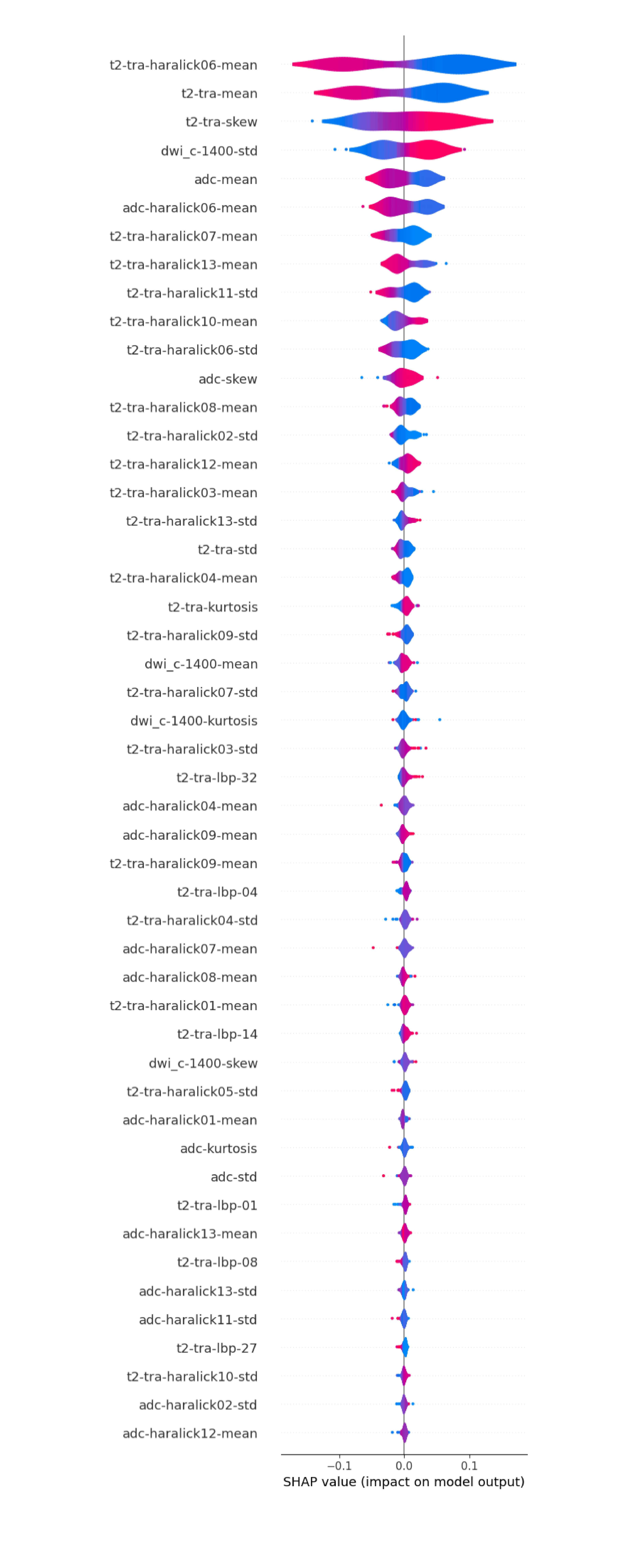}}
\caption{Shapley impact value distributions for best RF Classifier trained using all features without feature selection for both datasets. Only the 50 most significant features are shown. The remaining features are negligible. The color scale indicates feature values ranging from low (blue) to high (red).}
\end{figure}

In the previous section, we considered the impact of various features of the best-performing RF classifier with SBFS feature selection. We also trained an RF classifier of the same type using all features without SBFS. Fig.~\ref{fig:rfc-shapley-no-feature-selection} shows the Shapley decision trees, including the most important features for the best RF classifiers without feature selection for both datasets. Figs.~\ref{fig:rfc-shapley-no-feature-selection-px} and~\ref{fig:rfc-shapley-no-feature-selection-sw} show the Shapley distributions for the top-50 features for the Prostate-X and in-house datasets, respectively. Features omitted due to space constraints have negligible impact. Although slightly more features contribute, most features still have negligible impact; this holds in particular for LBP features. Features derived from all modalities contribute. First order statistical features still play a dominant role but more Haralick features contribute as well. Most importantly, using all features reduces performance of the algorithm compared to best RF algorithm with SBSF.

\section{Conclusions}

Shapley-value based feature analysis suggests that only a few features determine the classification outcome in most cases, with a dominant role played by first-order statistical features, and a limited number of Haralick texture features. Local binary patterns play no significant role for the best-performing RF classifiers for any of the datasets considered. However, the fact the best algorithms used features from all three modalities suggests that all are contributing valuable information. Also, there is no significant difference for different patch sizes (rescaled from the original size). The fact that the classification results are consistently determined by a small subset of features suggests that many features are redundant, and could be used to design streamlined classification algorithms using fewer features. Despite the limited generalizability of the results, the consistency observed in terms of the relevant features as well as the best classifiers types is encouraging. The much clearer decision trees for the in-house dataset compared to the far more complex split for the Prostate-X dataset require further exploration. It could indicate that classification of early-stage PCa may be clearer, but it may also indicate that larger datasets are needed to cover the covariance. Important next steps are to extend results to larger datasets focused on early-state PCa identification. Despite the traditional machine learning classifiers performing better than their deep learning counterparts, it may be interesting to see which features the deep learning methods use and how they compare to those identified in this analysis, especially as traditional machine learning classifiers may not perform as well on more complex classification tasks on larger datasets. The features identified may also form the basis for specifying specific textures in the various modalities and their relations to identify cancer and explain the decisions.

\section*{Acknowledgements}
We thank the staff of Swansea University's Clinical Imaging Unit, especially Superintendent Research Radiographer, Anthony Rees, and former clinical director, Dr Rhodri Evans, for providing and annotating the in-house dataset.

\FloatBarrier

% comment out these lines to use output.bbl file used in submission, does not appear to work with overleaf
%\bibliographystyle{splncs04}
%\bibliography{bibliography}

\end{document}